\documentstyle[aps,preprint]{revtex}
\begin{document}
\setcounter{page}{0}
\def\footnoterule{\kern-3pt \hrule width\hsize \kern3pt}
\tighten
\title{Large $N$ Limit of Higher Derivative Extended CP($N$) Model
}
\author{Taichi Itoh$^{a,b}$\footnote{Email address: taichi@knu.ac.kr} 
and Phillial Oh$^a$\footnote{
Email address: ploh@dirac.skku.ac.kr
}}
\address{
$^a$Department of Physics and Institute of Basic Science\\
Sungkyunkwan University, Suwon 440-746, Korea\\
$^b$Department of Physics, Kyungpook National University, 
Taegu 702-701 Korea}

\date{SKKUPT-01/2000, June 2000}

\maketitle
\draft 

\begin{abstract}
We construct a fourth-order derivative CP($N$) model in $1+1$ dimensions
by incorporating the topological charge density squared term into the
Lagrangian. We quantize the theory by reformulating with auxiliary fields
and then performing the path integral explicitly. We discuss the
renormalizability in the large $N$ limit and relevance of the effective
action with axion physics.  
\end{abstract}



\newpage
\noindent
1. Introduction.\\

The $1+1$ dimensional CP($N$) model has been studied
intensively in the large $N$ limit \cite{coleman}. It is known to be exactly 
soluble, and exhibits interesting phenomena such as dynamical mass generation
and asymptotic freedom which is also one of the essential properties of gauge
theories\cite{coleman}. Therefore it is important to extend the CP($N$)
model and investigate various properties. In this paper, we consider a higher
derivative extension of the model. More specifically, we invent a fourth-order
derivative interaction by utilizing the topological charge density on the
CP($N$). One of the merit of our higher derivative theory is that it permits
an auxiliary field formalism and the path integral can be performed
explicitly. The theory has only logarithmic divergence in the large $N$ limit, 
still it remains non-renormalizable due to the lack of counter terms. 
But we find that the effective theory is well-defined and describes a
two-dimensional axion interacting with Maxwell theory.

We start with the coadjoint orbit approach to nonlinear sigma model
\cite{oh} for some generality and introduce a
coadjoint orbit variable $Q =gKg^{-1}$ on $G/H$, where $g\in G$ and $K$ which
belongs to the Lie algebra ${\cal H}$ of $H$ is the centralizer for 
${\cal H}$. In order to construct higher derivative theory, we consider  
$t\equiv\epsilon^{\mu\nu}\!< Q\,\partial_\mu Q\,\partial_\nu Q>$ 
 which is the topological charge density on the coadjoint orbit
$G/H$. The topological charge $T=c_1\int\!d^2x\,t$ 
with some normalization constant $c_1$ is completely
characterized by the homotopy group $\pi_1(H)$. It is well known that this
group is in general given by (sum of) additive group of integers depending on
the nature of the centralizer $K$, and the 
topological charge $T$ characterizes the instanton solution
of the nonlinear sigma model described by the field $Q$ \cite{pere}.  
Our higher derivative model is constructed by adding the fourth derivative
term $t^2$ to the nonlinear sigma model : 
\begin{equation}
 S=\frac{1}{g^2} \int\!d^2x
\left(-\frac{1}{2}<\partial_\mu Q\,\partial^\mu Q> 
+\frac{t^2}{2M^2}\right),
\label{action}
\end{equation}
where $<\cdots>$ denotes trace 
and $M$ is some mass scale in the theory.
In SU(2) case, the above model reduces to the two-dimensional low energy
effective description of QCD which was proposed by 
Faddeev and Niemi \cite{fad}. 
Also, it is related with Skyrmion model, because
$t^2$ can be written as $t^2 \sim\,< J_{\mu\nu}J^{\mu\nu}>$,
$J_{\mu\nu}=[\partial_\mu Q, \partial_\nu Q]$.

Let us focus on CP($N$) case and study the large $N$ limit of
(\ref{action}) by using the path integral quantization method. We will show
that the above higher derivative theory admits an exact path integration
in the large $N$ analysis. In the CP($N$) orbit, the coadjoint variable $Q$
can be described by a single $N$ component complex column vector $z=(z_1,
\cdots, z_N)^T$ which defines the 
${\rm CP}(N)\equiv{\rm SU}(N)/{\rm SU}(N-1)\times{\rm U}(1)$
manifold  \cite{oh962}:
\begin{equation}
Q=-iz z^\dagger+i\frac{I}{N},\quad z^\dagger z=1.
\label{cpn}
\end{equation}
Then, the topological charge density is given by
\begin{equation}
\epsilon^{\mu\nu}\!<Q\,\partial_\mu Q\,\partial_\nu Q>\,=
i\epsilon^{\mu\nu} (\partial_\mu z)^\dagger (\partial_\nu z).
\label{topch}
\end{equation}
Note that this is quadratic in $z$ which is essential for the path integral in
the large $N$ limit. 

Expressing the Lagrangian (\ref{action}) in terms of $z^\dagger$ and $z$
and introducing two auxiliary fields $A_\mu(x)$ and $b(x)$, 
we obtain 
 \begin{equation}
{\cal L}=\frac{1}{g^2}\left[(D_\mu z)^\dagger (D^\mu z)
+i b\,\epsilon^{\mu\nu}(\partial_\mu z)^\dagger (\partial_\nu z)
-\frac{M^2}{2}\,b^2 -\lambda\,(z^\dagger z -1)\right],
\label{auxac}
\end{equation}
where $D_\mu \equiv \partial_\mu -iA_\mu$.
Note that $z$ with the constraint $z^\dagger z=1$ contains $2N-1$ real 
scalars, whereas the coset space is a ($2N-2$)-dimensional manifold. 
This mismatch is due to the local U(1) symmetry of the model 
which is obvious from (\ref{action}) and (\ref{cpn}):
$Q$ is invariant under the local U(1) phase rotation of $z$.
Specifically, the Lagrangian is invariant under the following U(1) gauge 
transformation:
\begin{equation}
z(x)\to e^{i\alpha(x)}z(x),\quad
A_\mu (x)\to A_\mu (x) +\partial_\mu \alpha(x),\quad
b(x)\to b(x),
\end{equation}
of which phase mode $\alpha(x)$ stands for a gauge redundancy in $z$.
Solving the equation of motion and eliminating $A_\mu$ and $b$ fields, 
we see that the extra second and third terms in the Lagrangian actually
provide the fourth derivative term $\frac{t^2}{2M^2}=
-\frac{1}{2M^2}\left[\epsilon^{\mu\nu}(\partial_\mu z)^\dagger 
(\partial_\nu z)\right]^2$
with the constraint $z^\dagger z=1$. Note that even in two dimensions this 
fourth derivative term is not renormalizable in perturbative sense. 
The leading divergence is given by a quadratic divergence in one-loop 
four-points functions. In the large $N$ limit, however, we find that
the only divergence is given by a logarithmic divergence.

In order to proceed, we separate the field $z$ 
into $2N-2$ Nambu-Goldstone bosons $\psi\equiv(z_1,\dots,z_{N-1})^T$
associated with the spontaneously broken SU($N$) symmetry and Higgs bosons 
$z_N \equiv \frac{g}{\sqrt{2}}(\sigma+i\chi)$. 
In general there are two possible phases: 
(I) $\langle\sigma\rangle\neq 0$, $\langle\lambda\rangle=0$ and
(II) $\langle\sigma\rangle=0$, $\langle\lambda\rangle\neq 0$.
In phase (I) both global SU($N$) and local U(1) symmetries are broken 
simultaneously and $\psi$ arise as massless Goldstone bosons. 
Through the Higgs mechanism $\chi$ turns to a longitudinal mode of 
massive gauge boson $A_\mu$. On the other hand in phase (II) both 
global SU($N$) and  local U(1) are not  spontaneously broken. 
Instead $\psi$ and 
$z_N$ are combined into $z$ with a universal mass 
$\langle\lambda\rangle^{1/2}$. 
Note that in two and less than two dimensions any continuous symmetry 
cannot be broken spontaneously (Coleman-Mermin-Wagner theorem) so that 
the phase (II) is the only possible phase in two dimensions.\\ 

\noindent
2. Large $N$ effective action\\

The $1/N$ expansion provides a systematic way of non-perturbative 
resummation of Feynman diagrams. Even if a given theory is 
non-renormalizable in perturbative expansion, it possibly can turn into a 
renormalizable theory through such a resummation technique as the $1/N$ 
expansion. In fact the CP($N$) model is non-renormalizable in larger 
than two dimensions, whereas the model in less than four dimensions 
can be renormalized in $1/N$ expansion. 

We can rewrite the Lagrangian up to total derivative terms as
\begin{equation}
{\cal L}=\frac{1}{g^2}\,z^\dagger\left[-\,\partial^2 -m^2
-\Gamma\right]z+\frac{\lambda}{g^2}-\frac{M^2}{2g^2}\,b^2,
\end{equation}
where we separate the Goldstone boson mass $m^2$ from 
$\lambda\equiv m^2 +\tilde{\lambda}$ and $\Gamma$ stands for the 
interaction terms:
\begin{equation}
\Gamma\equiv-iA_\mu (\partial^\mu -\overleftarrow{\partial^\mu})
-A_\mu A^\mu -ib\,\overleftarrow{\partial^\mu}\epsilon_{\mu\nu}
\partial^\nu,
\end{equation}
where $\overleftarrow{\partial^\mu}$ does not operate on both 
$A_\mu$ and $b$. The large $N$ effective action is given by
\begin{equation}
S_{\rm eff}
=\int\!d^2 x\,{\cal L}+iN\,{\rm Tr}\,{\rm Ln}\!
\left[-\,\partial^2 -m^2\right]-iN\sum_{n=1}^{\infty}
\frac{1}{n}\,{\rm Tr}\!\left[\frac{1}{-\,\partial^2 -m^2}\,
\Gamma\right]^n.
\end{equation}
After some straightforward calculations
with momentum cutoff $\Lambda$ in a gauge invariant way, 
the effective action
up to quadratic terms ($n=1,2$) is obtained as
\begin{eqnarray} 
S_{\rm eff} &=& N\int\!d^2 x\,\Biggl[\frac{1}{Ng^2}\, 
z^\dagger\left[-\,\partial^2 -m^2 -\Gamma\right]z 
-\frac{1}{2}\frac{M^2}{Ng^2}\,b^2 
\nonumber\\ && 
+(m^2 +\tilde{\lambda})\left[\frac{1}{Ng^2} 
-\frac{1}{4\pi}\ln\frac{\Lambda^2}{m^2}\right]-\frac{1}{4\pi}\,m^2 
+\frac{1}{2}\,\tilde{\lambda}\,\Pi_{\lambda}(i\partial)\,\tilde{\lambda} 
\nonumber\\ && 
-\frac{1}{4}\,F_{\mu\nu}\,\Pi_{A}(i\partial)\,F^{\mu\nu} 
+\frac{1}{2}\,b\,\Pi_{bA}(i\partial)\,\epsilon^{\mu\nu}F_{\mu\nu} 
-\frac{1}{4}\,b\,\partial^2 \Pi_{bA}(i\partial)\,b\,\Biggr]. 
\label{Seff} 
\end{eqnarray} 
where the vacuum polarization functions are given by
\begin{eqnarray}
\Pi_{\lambda}(p)&=&\frac{1}{2\pi p^2}\sqrt{\frac{-p^2}{4m^2 -p^2}}\,
\ln\!\left(\frac{\sqrt{4m^2 -p^2}-\sqrt{-p^2}}
{\sqrt{4m^2 -p^2}+\sqrt{-p^2}}\,\right),\\
\Pi_{A}(p)&=&\frac{1}{\pi p^2}+\left(1-\frac{4m^2}{p^2}\right)
\Pi_{\lambda}(p),\\
\Pi_{bA}(p)&=&-\frac{1}{4\pi}\ln\frac{\Lambda^2}{m^2}+\frac{1}{4\pi}
-\frac{1}{2}\,p^2\,\Pi_{A}(p).
\end{eqnarray}
Note that there arise logarithmic divergences in $b\,\partial^2 b$ and 
$b\,\epsilon^{\mu\nu}F_{\mu\nu}$ terms which have no counter terms 
in the original Lagrangian.\\

\noindent
3. Renormalization in $1/N$ leading order\\

Renormalization of the coupling $g$ can be worked out in the same manner 
as in the original CP($N$) model. The large $N$ effective potential 
is defined as the effective action divided by $\Omega\equiv\int d^2 x$ 
with $\tilde{\lambda}$, $A_\mu$, $b$, $z$, $z^\dagger$ all set equal to zero. 
It is given by
\begin{equation}
\frac{1}{N}V_{\rm eff}=-\frac{m^2}{N g^2}+\frac{m^2}{4\pi}
\left[\ln\frac{\Lambda^2}{m^2}+1\right],
\end{equation}
where $N g^2$ is fixed finite in the large $N$ limit so that the $1/N$ 
expansion can be treated systematically as a loop expansion. 
The Goldstone boson mass $m$ is determined as a nontrivial solution to the 
gap equation: 
\begin{equation}
\frac{dV_{\rm eff}}{dm^2}=0 \quad\longleftrightarrow\quad 
\frac{1}{g^2}=\frac{N}{4\pi}\ln\frac{\Lambda^2}{m^2},
\label{gapeqn}
\end{equation}
from which $m^2$ reads
\begin{equation}
m^2 =\Lambda^2 \exp\left[-\frac{4\pi}{N g^2}\right].
\end{equation}
We notice that $m$ can be independent of the ultraviolet (UV) cutoff 
$\Lambda$ by imposing $\Lambda$ dependence on the coupling $g$. 
In fact the scale invariance condition $\Lambda\,dm/d\Lambda=0$ leads us 
to the Callan-Symanzik $\beta$-function
\begin{equation}
\beta(g)\equiv\Lambda\frac{dg}{d\Lambda}=-\frac{N g^3}{4\pi},
\end{equation}
which shows the asymptotically free behavior of the coupling.
In the original CP($N$) the only divergence is the one which  arises in the
large  $N$ effective action through a tadpole diagram coupled with
$\tilde{\lambda}$  so that the scale invariance condition
$\Lambda\,dm/d\Lambda=0$ is enough  to achieve the cutoff independent theory.
Since $m$ is scale invariant,  the gap equation (\ref{gapeqn}) suggests the
renormalization of coupling  is given by
\begin{equation}
\frac{1}{Ng^2}-\frac{1}{4\pi}\ln\frac{\Lambda^2}{m^2}
=\frac{1}{Ng_R^2}-\frac{1}{4\pi}\ln\frac{\mu^2}{m^2},
\label{gr}
\end{equation}
where $g_R$ is the renormalized coupling at the reference energy 
scale $\mu$. 

In the extended model, however, logarithmic divergences arise in the 
coefficients of the induced extra terms, $b\,\partial^2 b$ and 
$b\,\epsilon^{\mu\nu}F_{\mu\nu}$ which do not have their 
counter terms in the classical action. Therefore our theory is 
not renormalizable though the coupling $g$ can be renormalized in the same 
way as in the original CP($N$) model. 
Let us look at how this argument works in the large $N$ effective 
action (\ref{Seff}). 
The induced kinetic terms of $A_\mu$ and $\tilde{\lambda}$ are UV finite 
in themselves so that we do not need wave function renormalization for them.
Then the third term in the right hand side of Eq.\ (\ref{Seff}) becomes UV 
finite through Eq.\ (\ref{gr}) from which the Z factor for the coupling can 
be read 
\begin{equation}
Z^{-1}\equiv\frac{g_R^2}{g^2}=1+\frac{Ng_R^2}{4\pi}
\ln\frac{\Lambda^2}{\mu^2}.
\end{equation}
The kinetic term of $z$ has to be UV finite in itself so that we see
\begin{equation}
\frac{1}{Ng^2}\,z^\dagger\left[-\,\partial^2 -m^2 \right]z=
\frac{1}{Ng_R^2}\,z_R^\dagger\left[-\,\partial^2 -m^2 \right]z_R,
\end{equation}
where $z_R$ has been introduced through $z=Z_z^{1/2}z_R$ and $Z_z$ is 
thereby determined as $Z_z \equiv Z$ in order to cancel the Z factor 
from the coupling renormalization. Thus we realize that $\Gamma$ in the 
effective action has to remain invariant through renormalization procedure. 
This forces all of $A_\mu$, $\tilde{\lambda}$ and $b$ to be unchanged 
through renormalization. Thus we obtain another renormalization condition
\begin{equation}
\frac{M^2}{Ng^2}=\frac{M_R^2}{Ng_R^2},
\end{equation}
which determines the Z factor for $M^2$ as $Z_M \equiv Z$.
On the other hand,  we do 
not have any degrees of freedom to subtract the logarithmic divergences 
which arise in the kinetic term of $b$ and in the mixing term between 
$b$ and $A_\mu$. This makes our extended model non-renormalizable 
even after the $1/N$ resummation.

Even though the theory is not renormalizable, the effective theory is a well
defined renormalizable theory. 
In fact, by using the gap equation (\ref{gapeqn}), 
the logarithmic $\Lambda$ dependence of $\Pi_{bA}(i\partial)$ 
in (\ref{Seff}) can be completely eliminated in 
favor of the $1/N$ counting rule $Ng^2 =\mbox{fixed}$, and we see that 
the effective action for $A_\mu$ and $b$ field is given by 
\begin{eqnarray}
S_{\rm eff} [A_\mu, b]=\frac{1}{g^2}\int\!d^2x\left[ 
\frac{c}{4}\,(\partial_\mu b)(\partial^\mu b) 
+\frac{c}{2}\,b\,\epsilon^{\mu\nu}F_{\mu\nu} 
-\frac{M^2}{2}\,b^2\right]-
\frac{N}{48\pi m^2}\int\!d^2x\, F_{\mu\nu}F^{\mu\nu},
\end{eqnarray}
where we have expanded the vacuum polarization functions 
with respect to $p^2 /4m^2$ after the analytic continuation from $p^2 <0$ 
to $0\le p^2 <4m^2$. Here, the constant $c\equiv Ng^2/4\pi -1$ confines 
the physical region of coupling to $g^2 >4\pi/N$. 
This is a two-dimensional model in which the Maxwell field $A_\mu$ interacts
with a pseudoscalar field $b$ with $b\,^*\!F$ axion type interaction
\cite{axion}.    We note that the potential prefers $\langle b \rangle=0$ for
the minimum, and the effect of  CP violating term  $\langle b\rangle\,^*\!F$
will be suppressed. 

The above argument has meaning only when we suppose that the
large $N$ effective gauge theory has instanton solutions \cite{abda}.
However, these solutions are usually destroyed by the next to leading order
corrections in $1/N$.  Recall also even though the original CP($N$) model has
instanton solutions \cite{pere},  the inclusion of our higher derivative term
does not admit any instanton solutions. Nevertheless,   the aforementioned
suppression mechanism may  provide a 
useful toy model for the strong CP problem \cite{poly}.\\

\noindent
4. Conclusion\\

In summary, we have constructed a higher derivative CP($N$) model and
quantized it by using the path integral method. We have illustrated that in
the large $N$ limit,   the ultraviolet divergence can be completely isolated
into a logarithmic divergence but the theory remains non-renormalizable due to
the lack of counter terms. However, we have found that the effective action is
a well-defined renormalizable theory, and it describes a massless gauge
field interacting with massive axion. It would be interesting to check
whether such a composite higher derivative axion model could have some
relevance in higher dimensional axion physics. \\

T.I. was supported by KOSEF Postdoctoral Fellowship and Korea Research
Center for Theoretical Physics and Chemistry.
P.O. was supported by the Samsung Research Fund, Sungkyunkwan
University, 1999. This work was also partially supported by BK21 Physics
Research Program.

\end{document}